\documentclass[aps,twocolumn,amsmath,amssymb,preprintnumbers]{revtex4}
\usepackage{amsmath} \usepackage{amsfonts} \usepackage{amssymb}
\usepackage{bbm}
\usepackage{epsfig}
\usepackage{graphics}
\usepackage{graphicx}
\newcommand{\be}{\begin{equation}}
\newcommand{\ee}{\end{equation}}
\newcommand{\ba}{\begin{eqnarray}}
\newcommand{\ea}{\end{eqnarray}}
\newcommand{\nn}{\nonumber}
\newcommand{\kr}{\rangle}
\newcommand{\kl}{\langle}
\newcommand{\M}{^{(M)}}
\newcommand{\hmn}{^{mn}}
\newcommand{\tmn}{_{mn}}

\newcommand{\x}{\tilde x_}

\newcommand{\f}{{\cal F}}
\newcommand{\D}{{\cal D}}
\newcommand{\g}{{\cal G}}
\newcommand{\h}{{\cal H}}

\begin{document}

\title[ ]{Lattice spinor gravity}

\author{C. Wetterich}
\affiliation{Institut  f\"ur Theoretische Physik\\
Universit\"at Heidelberg\\
Philosophenweg 16, D-69120 Heidelberg}

\begin{abstract}
We propose a regularized lattice model for quantum gravity purely formulated
in terms of fermions. The lattice action exhibits local Lorentz symmetry,
and the continuum limit is invariant under general coordinate
transformations. The metric arises as a composite field. Our lattice model
involves no signature for space and time, describing simultaneously a
Minkowski or euclidean theory. It is invariant both under Lorentz
transformations and euclidean rotations. The difference between space and
time arises from expectation values of composite fields. Our formulation
includes local gauge symmetries beyond the generalized Lorentz symmetry. The
lattice construction can be employed for formulating models with local gauge
symmetries purely in terms of fermions.
\end{abstract}

\maketitle

It has often been advocated that there could be basic incompatibilities between quantum mechanics and general relativity. In contrast, a different line of thought maintains that gravity may be formulated according to the same principles as any other quantum field theory. While Einstein's gravity is not perturbatively renormalizable, a nontrivial ultraviolet fixed point could permit non-perturbative renormalizability. This scenario of ``asymptotic safety'' \cite{Wei} has found support by recent investigations \cite{Reu,Cod} using functional renormalization based on the effective average action or flowing action \cite{CWRG}. In this case one may conjecture that quantum gravity can ultimately be defined by a suitable functional integral. Such a well defined regularized functional integral, similar to lattice gauge theories, is still missing. Several proposals have encountered various obstacles, mainly from the difficulty to implement diffeomorphism symmetry if the metric is used as the basic degree of freedom. In this letter we propose a functional integral for quantum gravity based on Grassmann variables. Our lattice formulation is well defined for a finite number of lattice sites. The continuum limit is obtained as usual by decreasing the lattice distance at fixed physical length scale. (For alternative lattice approaches, e.g. based on dynamical triangulation, see ref. \cite{Am}, \cite{Ha}.)

For any lattice regularization of gravity it is crucial that the symmetry of diffeomorphisms (general coordinate invariance) is realized for the continuum limit. This will guarantee the presence of a massless graviton. If any kind of derivative expansion is possible for distances much larger than the lattice distance, one further expects an effective action for the graviton that is dominated by the Einstein-Hilbert action, possibly with a cosmological constant. If fermions are present, as in our approach, another crucial property is local Lorentz symmetry.

We formulate a model based purely on spinors - spinor gravity. The basic degrees of freedom are fermions and the functional integral is based on Grassmann variables. The geometrical degrees of freedom as the metric and the vierbein arise as expectation values of bosonic composite fields. For continuous spacetime the action of spinor gravity is diffeomorphism invariant and can indeed contain Einstein's curvature scalar, as shown in a loop expansion for a similar theory \cite{HCW,CWSG}. Early formulations of spinor gravity as in \cite{HCW,CWSG} exhibit, however, only global and not local Lorentz symmetry. This typically induces additional torsion invariants in the gravitational effective action. The issue of global instead of local Lorentz symmetry has been extensively discussed in \cite{CWSG}. It was found that one of the torsion invariants - the only one generated at one loop order - is actually compatible with all present observations, while a second possible invariant is excluded by the tests of general relativity. In this letter we avoid this difficulty by formulating a model with local Lorentz symmetry, with analogies to the higher dimensional model in ref. \cite{CWLL}. First observations that a diffeomorphism invariant action for fermions can be formulated without the use of a metric, and the conjecture that the metric is a composite field, have been made long ago \cite{Aka,Ama,Den}.(The actual implementation in these approaches is not fully consistent - for example the inverse of products of Grassmann variables does not exist.) We build on these ideas, but we propose a different action that implements local Lorentz symmetry.

In this letter we ask the simple question if a lattice formulation of spinor gravity is possible which obeys the following four criteria: (1) For a finite number of lattice points the functional integral should be mathematically well defined. (2) The lattice action should be invariant under local Lorentz transformations. (3) A continuum limit should exist where gravitational interactions remain present at distances large compared to the lattice distance. (4) The continuum limit of the action should be diffeomorphism invariant, and there should be a lattice origin of this symmetry. 

The answer to this question is positive. In constructing such a lattice model, we find additional symmetries. For two flavors of fermions the continuum limit exhibits an $SU(2)_L\times SU(2)_R$ gauge symmetry. Most important, the local Lorentz-transformations of the group $SO(1,3)$ are extended to complex transformation parameters realizing the group $SO(4,{\mathbbm C})$, which also includes the euclidean rotation symmetry $SO(4)$. No signature for space and time are singled out in the basic formulation - both appear on completely equal footing. The difference in signature between space and time arises as a dynamical effect through expectation values of composite fields \cite{CWLL}.

\newpage\noindent
{\bf 1. Action and functional integral}

Let us explore a setting with $16$ Grassmann variables $\psi^a_\gamma$ at every spacetime point $x$, $\gamma=1\dots 8,~a=1,2$. The coordinates $x$ parametrize the four dimensional vector space or real numbers ${\mathbbm R}^4$, i.e. $x^\mu=(x^0,x^1,x^2,x^3)$. We will later associate $t=x^0$ with a time coordinate, and $x^k,~k=1,2,3$, with space coordinates. There is, however, a priori no difference between time and space coordinates. We will work with complex Grassmann variables $\varphi^a_\alpha,\alpha=1\dots 4$, 
\be\label{AA1}
\varphi^a_\alpha(x)=\psi^a_\alpha(x)+i\psi^a_{\alpha+4}(x),
\ee
with $\alpha$ the ``Dirac index'' and $a$ the ``flavor index''. We propose an action which involves twelve Grassmann variables and realizes diffeomorphism symmetry and local $SO(4, {\mathbbm C})$ symmetry 
\ba\label{A1}
S&=&\alpha\int d^4x\varphi^{a_1}_{\alpha_1}\dots\varphi^{a_8}_{\alpha_8}
\epsilon^{\mu_1\mu_2\mu_3\mu_4}\\
&&\times J^{a_1\dots a_8b_1\dots b_4}_{\alpha_1\dots\alpha_8\beta_1\dots\beta_4}
\partial_{\mu_1}\varphi^{b_1}_{\beta_1}\partial_{\mu_2}\varphi_{\beta_2}^{b_2}\partial_{\mu_3}
\varphi_{\beta_3}^{b_3}\partial_{\mu_4}\varphi^{b_4}_{\beta_4}+c.c.,\nn
\ea
where we sum over repeated indices. The complex conjugation $c.c.$ replaces $\alpha\to\alpha^*,J\to J^*$ and $\varphi_\alpha(x)\to\varphi^*_\alpha(x)=\psi_\alpha(x)-i\psi_{\alpha+4}(x)$, such that $S^*=S$. In terms of the Grassmann variables $\psi^a_\gamma(x)$ the action $S$ as well as $\exp(-S)$ are elements of a real Grassmann algebra. 

Invariance of the action under general coordinate transformations follows from the use of the totally antisymmetric product of four derivatives $\partial_\mu=\partial/\partial x^\mu$. Indeed, with respect to diffeomorphisms $\varphi(x)$ transforms as a scalar, and $\partial_\mu\varphi(x)$ as a vector. The particular contraction with the totally antisymmetric tensor $\epsilon^{\mu_1\mu_2\mu_3\mu_4},\epsilon^{0123}=1$, allows for a realization of diffeomorphism symmetry without the use of a metric. 

The partition function $Z$ is defined as
\ba\label{1A}
Z&=&\int {\cal D}\psi g_f\exp (-S)g_{in},\nn\\
\int {\cal D}\psi&=&\prod_x \prod^{2}_{a=1}
\big\{ \int d\psi^a_1(x)\dots \int d\psi^a_8(x)\big\}.
\ea
Later we will use discrete spacetime points on a lattice such that the Grassmann functional integral \eqref{1A} is well defined mathematically. We assume that the time coordinate $x^0=t$ obeys $t_{in}\leq t\leq t_f$. The boundary term $g_{in}$ is a Grassmann element constructed from $\psi_\gamma(t_{in},\vec x)$, while $g_f$ involves terms with powers of $\psi_\gamma(t_f,\vec x)$, were $\vec x=(x^1,x^2,x^3)$. If $g_{in}$ and $g_f$ are elements of a real Grassmann algebra the partition function is real. We may restrict the range of the space coordinates or use a torus $T^3$ instead of ${\mathbbm R}^3$. For a discrete spacetime lattice the number of Grassmann variables is then finite. 

Observables ${\cal A}$ will be represented as Grassmann elements constructed from $\psi_\gamma(x)$. We will consider only bosonic observables that involve an even number of Grassmann variables. Their expectation value is defined as
\be\label{2A}
\kl {\cal A}\kr=Z^{-1}\int{\cal D}\psi g_f{\cal A}\exp (-S)g_{in}.
\ee
``Real observables'' are elements of a real Grassmann algebra, i.e. they are sums of powers of $\psi_\gamma(x)$ with real coefficients. For real $g_{in}$ and $g_f$ all real observables have real expectation values. We will take the continuum limit of vanishing lattice distance at the end. Physical observables are those that have a finite continuum limit. 

\medskip\noindent
{\bf 2. Generalized Lorentz transformations}

We first require the action to be invariant under global generalized Lorentz transformations. Thus the tensor $J^{a_1\dots a_8b_1\dots b_4}_{\alpha_1\dots \alpha_8\beta_1\dots\beta_4}$ must be invariant under global $SO(4,{\mathbbm C})$ transformations. We will often use double indices $\epsilon=(\alpha,a)$ or $\eta=(\beta,b)$, $\epsilon, \eta=1\dots 8$. The tensor $J_{\epsilon_1\dots \epsilon_8\eta_1\dots\eta_4}$ is totally antisymmetric in the first eight indices $\epsilon_1\dots \epsilon_8$, and totally symmetric in the last four indices $\eta_1\dots\eta_4$. This follows from the anticommuting properties of the Grassmann variables $\varphi_\epsilon\varphi_\eta=-\varphi_\eta\varphi_\epsilon$. We will see that for an invariant $J$ the action \eqref{A1} is also invariant under local $SO(4, {\mathbbm C})$ transformations.

Local $SO(4,{\mathbbm C})$ transformations act infinitesimally as
\be\label{A2}
\delta\varphi^a_\alpha(x)=-\frac12\epsilon_{mn}(x)(\Sigma^{mn}_E)_{\alpha\beta}\varphi^a_\beta(x),
\ee
with arbitrary complex parameters $\epsilon_{mn}(x)=-\epsilon_{nm}(x),m=0,1,2,3$. The complex $4\times 4$ matrices $\Sigma^{mn}_E$ are associated to the generators of $SO(4)$ in the (reducible) four-component spinor representation. They can be obtained from the euclidean Dirac matrices
\ba\label{A3}
\Sigma^{mn}_E=-\frac14[\gamma^m_E,\gamma^n_E]~,~\{\gamma^m_E,\gamma^n_E\}=2\delta^{mn}.
\ea
Subgroups of $SO(4,{\mathbbm C})$ with different signatures obtain by appropriate choices of $\epsilon_{mn}$. Real parameters  $\epsilon_{mn}$ correspond to euclidean rotations $SO(4)$. Taking $\epsilon_{kl}, k,l=1,2,3$ real, and $\epsilon_{0k}=-i\epsilon^{(M)}_{0k}$ with real $\epsilon^{(M)}_{0k}$, realizes the Lorentz transformations $SO(1,3)$. The Lorentz transformations can be written equivalently with six real transformation parameters $\epsilon^{(M)}_{mn}~,~\epsilon\M_{kl}=\epsilon_{kl}$, using Lorentz-generators $\Sigma\hmn_M$ and signature $\eta\hmn=diag(-1,1,1,1)$, 
\be\label{3A}
\delta\varphi=-\frac12\epsilon\M\tmn\Sigma\hmn_M\varphi ,
\ee
with
\be\label{A4}
\Sigma\hmn_M=-\frac14[\gamma^m_M,\gamma^n_M]~,~\{\gamma^m_M,\gamma^n_M\}=\eta^{mn}.
\ee
The euclidean and Minkowski Dirac matrices are related by $\gamma^0_M=-i\gamma^0_E,\gamma^k_M=\gamma^k_E$. 

The transformation of a derivative involves an inhomogeneous part 
\be\label{A5}
\delta\partial_\mu\varphi_\beta=-\frac12\epsilon\tmn(\Sigma\hmn\partial_\mu
\varphi)_\beta-\frac12\partial_\mu\epsilon\tmn(\Sigma\hmn\varphi)_\beta,
\ee
with $\Sigma\hmn=\Sigma\hmn_E~,~\gamma^m=\gamma^m_E$. The first ``homogeneous term'' $\sim \partial_\mu\varphi$ transforms as $\varphi_\beta$. Thus an invariant tensor $J$ guarantees an invariant action if the second term in eq. \eqref{A5} can be neglected. Contributions of the second ``inhomogeneous term'' to the  variation of the action $\delta S$ involve at least nine spinors at the same position $x$, i.e. $(\Sigma\hmn\varphi)^b_\beta(x)\varphi^{a_1}_{\alpha_1}(x)\dots\varphi^{a_8}_{\alpha_8}(x)$. Therefore this inhomogeneous contribution to $\delta S$ vanishes due to the identity $\varphi_\alpha(x)\varphi_\alpha(x)=0$ (no sum here) - at most eight different complex spinors can be placed on a given position $x$. The invariance of $S$ under global $SO(4,{\mathbbm C})$ transformations entails the invariance under local $SO(4,{\mathbbm C})$ transformations. We have constructed in ref. \cite{CWLL} a model for sixteen dimensional spinor gravity with local $SO(16,{\mathbbm C})$ symmetry. The present four-dimensional model shows analogies to this. 

It is important that all invariants appearing in the action \eqref{A1} involve either only factors of $\varphi_\alpha=\psi_\alpha+i\psi_{\alpha+4}$ or only factors of $\varphi^*_\alpha =\psi_\alpha-i\psi_{\alpha+4}$. It is possible to construct $SO(1,3)$ invariants which involve both $\varphi$ and $\varphi^*$. Those will not be invariant under $SO(4,{\mathbbm C})$, however. We can also construct invariants involving $\varphi$ and $\varphi^*$ which are invariant under euclidean $SO(4)$ rotations. They will not be invariant under $SO(1,3)$. The only terms which are invariant under both $SO(4)$ {\em and} $SO(1,3)$, and more generally $SO(4,{\mathbbm C})$, are those constructed from $\varphi$ alone or $\varphi^*$ alone, or products of such invariants. (Invariants involving both $\varphi$ and $\varphi^*$ can be constructed as products of invariants involving only $\varphi$ with invariants involving only $\varphi^*$.) 

We conclude that for a suitable $SO(4)$-invariant tensor $J$ the action has the symmetries required for a realistic theory of gravity for fermions, namely diffeomorphism symmetry and local $SO(1,3)$ Lorentz symmetry. No signature and no metric are introduced at this stage, such that there is no difference between time and space \cite{CWLL}. This follows from the fact that for an action of the type (2) local $SO(4,{\mathbbm C})$ symmetry is realized for every invariant tensor $J$. Here we define the $SO(4,{\mathbbm C})$-variation of arbitrary tensors with Dirac indices $\alpha_1\dots \alpha_N$ as
\be\label{SA}
\delta T_{\alpha_1\dots\alpha_N}=T_{\tilde\alpha\alpha_2\dots\alpha_N}\Sigma_{\tilde\alpha\alpha_1}+\dots
+T_{\alpha_1\dots\tilde\alpha}\Sigma_{\tilde\alpha\alpha_N},
\ee
with 
\be\label{SB}
\Sigma_{\alpha\beta}=-\frac12\epsilon_{mn}\Sigma^{mn}_{\alpha\beta}.
\ee
We can express global $SO(4,{\mathbbm C})$-transformations (with $\epsilon_{mn}$ independent of $x$) of the action equivalently by a transformation \eqref{A2} of the spinors $\varphi$ with fixed $J$, or by a transformation \eqref{SA} of $J$ with fixed $\varphi$. For $\delta J=0$ the action is invariant under global $SO(4,{\mathbbm C})$-transformations.

\medskip\noindent
{\bf 3. Weyl spinors}

Our model with two flavors allows us to construct symmetric invariants with two Dirac indices
\be\label{W3a}
S^\pm_{\eta_1\eta_2}=(S^\pm)^{b_1b_2}_{\beta_1\beta_2}
=\mp(C_\pm)_{\beta_1\beta_2}(\tau_2)^{b_1b_2}
\ee
with Pauli matrices $\tau_k$. The invariant tensors $C_\pm$ are antisymmetric
\be\label{SD}
(C_\pm)_{\beta_2\beta_1}=-(C_\pm)_{\beta_1\beta_2},
\ee
such that $S^\pm$ is symmetric under the exchange $(\beta_1,b_1)\leftrightarrow(\beta_2,b_2)$, or, in terms of the double index $\eta=(\beta,b)$, 
\be\label{SE}
S^\pm_{\eta_2\eta_1}=S^\pm_{\eta_1\eta_2}.
\ee

The $SO(4,{\mathbbm C})$-invariants $C_\pm$ can best be understood in terms of Weyl spinors. The matrix 
\be\label{37A}
\bar\gamma=-\gamma^0\gamma^1\gamma^2\gamma^3
\ee
commutes with $\Sigma\hmn$ such that the two doublets
\be\label{A25}
\varphi_+=\frac12(1+\bar\gamma)\varphi~,~\varphi_-=\frac12(1-\bar\gamma)\varphi
\ee
correspond to inequivalent two component complex spinor representations (Weyl spinors). We employ here a representation of the Dirac matrices $\gamma^m$ where $\bar\gamma=diag(1,1,-1,-1)$, namely
\be\label{16A}
\gamma^0=\tau_1\otimes 1~,~\gamma^k=\tau_2\otimes \tau_k.
\ee
(The general structure is independent of this choice. Our representation corresponds to the Weyl basis of ref. \cite{CWMS} where details of conventions can be found.) In this representation one has $\varphi^a_+=(\varphi^a_1,\varphi^a_2),\varphi^a_-=(\varphi^a_3,\varphi^a_4)$. We may order the double index $\eta$ or $\epsilon$ such that 
\ba\label{W1}
\varphi_{+,\eta}&=&(\varphi_1,\varphi_2,\varphi_3,\varphi_4)=(\varphi^1_1,\varphi^2_1,\varphi^1_2,\varphi^2_2)\nn\\
\varphi_{-,\eta}&=&(\varphi_5,\varphi_6,\varphi_7,\varphi_8)=(\varphi^1_3,\varphi^2_3,\varphi^1_4,\varphi^2_4),
\ea
i.e. $\beta=1,b=2$ corresponds to $\eta=2$.

An invariant matrix $C$ obeys
\be\label{A6H}
\Sigma^TC+C\Sigma=0.
\ee
In four dimensions, the matrix $C$ is antisymmetric \cite{CWS,CWMS}. There exist two matrices $C_1$ and $C_2$ which obey the condition \eqref{A6H} or
\be\label{A8}
C\Sigma\hmn C^{-1}=-(\Sigma\hmn)^T.
\ee
We can choose $C=C_1$ such that 
\be\label{A9}
C_1\gamma^mC^{-1}_1=-(\gamma^m)^T~,~C^T_1=-C_1~,~C^\dagger_1C_1=1,
\ee
and $C_1\gamma^m$  is a symmetric matrix
\be\label{A10}
(C_1\gamma^m)^T=C_1\gamma^m.
\ee
Another possible choice for $C$ obeying eq. \eqref{A8} is the antisymmetric matrix $C_2=C_1\bar\gamma$ which obeys
\be\label{A12}
C_2\gamma^mC^{-1}_2=(\gamma^m)^T~,~C^T_2=-C_2~,~(C_2\gamma^m)^T=-C_2\gamma^m.
\ee
The bilinears $\varphi C_1\varphi$ and $\varphi C_2\varphi$ correspond to the two singlets contained in the antisymmetric product of two Dirac spinors. In our basis one has $C_1=$diag($\tau_2,-\tau_2),~C_2=$diag$(\tau_2,\tau_2)$ and we introduce
\ba\label{W2}
C_+&=&\frac12(C_1+C_2)=\frac12 C_1(1+\bar\gamma)=
\left(\begin{array}{ccc}\tau_2&,&0\\0&,&0\end{array}\right),\nn\\
C_-&=&\frac12(C_1-C_2)=\frac12 C_1(1-\bar\gamma)=
\left(\begin{array}{ccc}0&,&0\\0&,&-\tau_2\end{array}\right).
\ea
such that $\psi^T_\pm C_1=\psi^T_\pm C_\pm=\psi^TC_\pm$.

It is straightforward to construct invariants only involving the two Weyl spinors $\varphi^1_+$ and $\varphi^2_+$. For this purpose we can restrict the index $\eta$ to the values $1\dots 4$. The action of $SO(4,{\mathbbm C})$ on $\varphi_+$ is given by the subgroup of complexified $SU(2,{\mathbbm C})_+$ transformations. In our basis the generators of $SU(2,{\mathbbm C})_+$ read
\be\label{W6}
\Sigma^{0k}=-\frac i2\tau_k~,~\Sigma^{kl}=\epsilon^{klm}\Sigma^{0m},
\ee
such that $\Sigma^{kl}$ is linearly dependent on $\Sigma^{0k}$. (For $SU(2,{\mathbbm C})_-$ the generators $\Sigma^{kl}$ are identical, while $\Sigma^{0k}=\frac i2\tau_k$. The subgroup of unitary transformations $SU(2)$ obtains for real transformation parameters, while we consider here arbitrary complex transformation parameters.) 

We observe that we can also consider a group $SU(2,{\mathbbm C})_L$ acting on the flavor indices of $\varphi_+$. With respect to $SU(2,{\mathbbm C})_+\times SU(2,{\mathbbm C})_L$ the four component spinor $\varphi_{+,\eta}~(\eta=1\dots 4)$ transforms as the $(2,2)$ representation. Since the matrix $(\tau_2)^{ab}$ in eq. \eqref{W3a} is invariant under $SU(2,{\mathbbm C})_L$, the invariant $S^+$ is invariant under the group 
\be\label{W7}
SO(4,{\mathbbm C})_+\equiv SU (2,{\mathbbm C})_+\times SU(2,{\mathbbm C})_L. 
\ee
(Here $SO(4,{\mathbbm C})_+$ should be distinguished from the generalized Lorentz transformation since it acts both in the space of Dirac and flavor indices.) With respect to $SO(4,{\mathbbm C})_+$ the two-flavored spinor $\varphi_+$ transforms as a four component vector. The classification of tensors, invariants and symmetries can be directly inferred from the analysis of four-dimensional vectors. Invariants only involving $\varphi_-$ can be constructed in a similar way with $SU(2,{\mathbbm C})_R$ acting on the flavor indices of $\varphi_-$ and $SO(4,{\mathbbm C})_-=SU(2,{\mathbbm C})_-\times SU(2,{\mathbbm C})_R$.

\medskip\noindent
{\bf 4. Action with local Lorentz symmetry}

A totally symmetric four index invariant can be constructed as
\ba\label{W9}
&&L_{\eta_1\eta_2\eta_3\eta_4}=\frac16
(S^+_{\eta_1\eta_2}S^-_{\eta_3\eta_4}+S^+_{\eta_1\eta_3}
S^-_{\eta_2\eta_4}+S^+_{\eta_1\eta_4}S^-_{\eta_2\eta_3}\nn\\
&&\hspace{1.0cm} +S^+_{\eta_3\eta_4}S^-_{\eta_1\eta_2}+S^+_{\eta_2\eta_4}
S^-_{\eta_1\eta_3}+S^+_{\eta_2\eta_3}S^-_{\eta_1\eta_4}).
\ea
The global invariant 
\be\label{30D}
D=\epsilon^{\mu_1\mu_2\mu_3\mu_4}
\partial_{\mu_1}\varphi_{\eta_1}\partial_{\mu_2}\varphi_{\eta_2}\partial_{\mu_3}\varphi_{\eta_3}\partial_{\mu_4}\varphi_{\eta_4}
L_{\eta_1\eta_2\eta_3\eta_4}
\ee
involves two Weyl spinors $\varphi_+$ and two Weyl spinors $\varphi_-$. Furthermore, an invariant with eight factors of $\varphi$ involves the totally antisymmetric tensor for the eight values of the double-index $\epsilon$
\ba\label{30E}
A^{(8)}&=&\frac{1}{8!}\epsilon_{\epsilon_1\epsilon_2\dots\epsilon_8}
\varphi_{\epsilon_1}\dots\varphi_{\epsilon_8}\nn\\
&=&\frac{1}{(24)^2}\epsilon_{\alpha_1\alpha_2\alpha_3\alpha_4}
\varphi^1_{\alpha_1}\dots\varphi^1_{\alpha_4}\epsilon_{\beta_1\beta_2\beta_3\beta_4}
\varphi^2_{\beta_1}\dots\varphi^2_{\beta_4}\nn\\
&=&\varphi^1_1\varphi^1_2\varphi^1_3\varphi^1_4
\varphi^2_1\varphi^2_2\varphi^2_3\varphi^2_4.
\ea

An action with local $SO(4,{\mathbbm C})$ symmetry takes the form 
\be\label{30F}
S=\alpha\int d^4xA^{(8)}D+c.c.
\ee
Indeed, the inhomogeneous contribution \eqref{A5} to the variation of $D(x)$ contains factors $(\Sigma^{mn}\varphi^b)_\beta(x)$. As discussed before, it vanishes when multiplied with $A^{(8)}(x)$, since the Pauli principle $\big(\varphi^a_\alpha(x)\big)^2=0$ admits at most eight factors $\varphi$ for a given $x$. In consequence, the inhomogeneous variation of the action \eqref{30F} vanishes and $S$ is invariant under {\em local } $SO(4,{\mathbbm C})$ transformations. In contrast to $\int d^4xD(x)$ the action $S$ is not a total derivative. Besides local $SO(4,{\mathbbm C})$, it is also invariant under local $SO(4,{\mathbbm C})_F$ gauge transformations, with $SO(4,{\mathbbm C})_F=SU(2,{\mathbbm C})_L\times SU(2,{\mathbbm C})_R$. 

The derivative-invariant $D$ can be written in the form 
\be\label{32A}
D=\epsilon^{\mu_1\mu_2\mu_3\mu_4}D^+_{\mu_1\mu_2}D^-_{\mu_3\mu_4},
\ee
with 
\be\label{32B}
D^\pm_{\mu_1\mu_2}=\partial_{\mu_1}\varphi_{\eta_1}S^\pm_{\eta_1\eta_2}\partial_{\mu_2}\varphi_{\eta_2}.
\ee
This shows that $D$ is invariant under the exchange $\varphi_{+,\eta}\leftrightarrow\varphi_{-,\eta}$. The transformation $\varphi\to\gamma^0\varphi$ maps $S^+_{\eta_1\eta_2}\leftrightarrow S^-_{\eta_1\eta_2}$ and therefore $D^+_{\mu_1\mu_2}\leftrightarrow D^-_{\mu_1\mu_2}$, such that again $D$ is invariant. (For our choice $\gamma^0=\tau_1\otimes 1$ the transformation $\varphi\to\gamma^0\varphi$ actually corresponds to $\varphi_{+,\eta}\leftrightarrow \varphi_{-,\eta}$.) We can also decompose
\be\label{32C}
A^{(8)}=A^+A^-,
\ee
with 
\be\label{32D}
A^+=\varphi^1_{+1}\varphi^1_{+2}\varphi^2_{+1}\varphi^2_{+2},
\ee
and similarly for $A^-$. The combinations 
\be\label{32E}
F^\pm_{\mu_1\mu_2}=A^\pm D^\pm_{\mu_1\mu_2}
\ee
are invariant under local $SO(4,{\mathbbm C})\times SO(4,{\mathbbm C})_F$ transformations. They involve six Weyl spinors $\varphi_+$ or six Weyl spinors $\varphi_-$, respectively. The action involves products of $F^+$ and $F^-$, 
\be\label{32F}
S=\alpha\int d^4x\epsilon^{\mu_1\mu_2\mu_3\mu_4}F^+_{\mu_1\mu_2}F^-_{\mu_3\mu_4}+c.c.
\ee

We define the Minkowski action by
\be\label{A23}
S=-iS_M~,~e^{-S}=e^{iS_M},
\ee
which yields the usual ``phase factor'' for the functional integral written in terms of $S_M$. We can define the operation of a transposition as a total reordering of all Grassmann variables. The result of transposition for a product of Grassmann variables depends only on the number of factors $N_\varphi$. For $N_\varphi=2,3$ mod $4$ the transposition results in a minus sign, while for $N_\varphi=4,5$ mod $4$ the product is invariant. In consequence, one finds that $S_M$ is symmetric. With respect to the complex conjugation c.c. used in eq. \eqref{A1} the Minkowski action is antihermitean. This complex conjugation, which is defined for the Grassmann variables $\psi_\gamma$ by the involution $\psi^a_{\alpha+4}\to-\psi^a_{\alpha+4}$ for $\alpha=1\dots 4$, is, however, not unique. We may define a different conjugation by an involution where the Grassmann variables changing sign are $\psi^1_5,\psi^1_6,\psi^1_7,\psi^1_8,\psi^2_3,\psi^2_4,\psi^2_5$ and $\psi^2_6$. In this case we use the same definition as before for $\varphi^1_\alpha$ and $\varphi^2_1,\varphi^2_2$, but we replace $\varphi^2_3$ and $\varphi^2_4$ by new complex Grassmann variables
\ba\label{AxA}
\xi^2_3&=&\psi^2_7-i\psi^2_3~,~\xi^2_4=\psi^2_8-i\psi^2_4,\nn\\
(\xi^2_3)^*&=&\psi^2_7+i\psi^2_3~,~(\xi^2_4)^*=\psi^2_8+i\psi^2_4.
\ea
The new complex conjugation can be interpreted as a multiplication of c.c. in eq. \eqref{A1} with the transformation $\varphi^2_-\to-\varphi^2_-$. Expanding the euclidean action in terms of $\varphi^1_\pm,\varphi^1_+$ and $\xi^2_-$ it changes sign under the new complex conjugation. With respect to this conjugation the Minkowski action is real and symmetric and therefore hermitean. We can use the first complex conjugation in order to establish that we work with a real Grassmann algebra, and the second one to define hermiticity of $S_M$ which is related to a unitary time evolution. 

\medskip\noindent
{\bf 5. Symmetries}

Besides the generalized Lorentz transformations $SO(4,{\mathbbm C})$ the action \eqref{30F}, \eqref{A18} is also invariant under continuous gauge transformations. By the same argument as for local $SO(4,{\mathbbm C})$ symmetry, any global continuous symmetry of the action is also a local symmetry due to the Pauli principle. We have already encountered the symmetry $SU(2,{\mathbbm C})_L$ which transforms
\be\label{S1}
\delta\varphi^a_{+\alpha}(x)=\frac i2\tilde\alpha_{+k}(x)(\tau_k)^{ab}\varphi^b_{+\alpha}(x),
\ee
with three complex parameters $\tilde\alpha_{+k}$, and similar for $SU(2,{\mathbbm C})_R$ acting on $\varphi_-$. For real $\tilde\alpha_{+k}$ these are standard gauge transformations with compact gauge group $SU(2)$. Altogether, we have four $SU(2,{\mathbbm C})$ factors, and with respect to $G=SU(2,{\mathbbm C})_+\times SU(2,{\mathbbm C})_-\times SU(2,{\mathbbm C})_L\times SU(2,{\mathbbm C})_R$ the Weyl spinors $\varphi_+$ and $\varphi_-$ transform as $(2,1,2,1)$ and $(1,2,1,2)$, respectively.

Discrete symmetries are a useful tool to characterize the properties of the model. Simple symmetries are $Z_{12}$ phase-transformations or multiplications with $\bar\gamma$ or $\gamma^0$, e.g.
\be\label{51}
\varphi\to \exp (2\pi in/12)\varphi~,~\varphi\to\bar\gamma\varphi~,~\varphi\to \gamma^0\varphi.
\ee
The reflection of the three space coordinates
\be\label{52}
\psi^a_\gamma(x)\to \psi^a_\gamma(Px)~,~P(x^0,x^1,x^2,x^3)=(x^0,-x^1,-x^2,-x^3),
\ee
changes the sign of the action. If this transformation is accompanied by any other discrete transformation which inverts the sign of $S$ the combined transformation amounts to a type of parity symmetry. As an example, we may consider
\be\label{53}
\varphi^1(x)\to\gamma^0\varphi^1(x)~,~\varphi^2(x)\to\gamma^0\bar\gamma\varphi^2(x).
\ee
Parity transformations can be constructed by combining the transformations \eqref{52} and \eqref{53}, together with some transformation that leaves $S$ invariant. For example, the transformation 
\be\label{54}
\varphi^1(x)\to\gamma^0_M\bar\gamma\varphi^1(Px)~,~\varphi^2(x)\to\gamma^0_M\varphi^2(Px)
\ee
leaves the action invariant. 

Time reflection symmetry can be obtained in a similar way by combining $\psi^a_\gamma(x)\to\psi^a_\gamma(-Px)$ with a suitable transformation that changes the sign of $S$, as for eq. \eqref{53}. Reflections of an even number of coordinates, including the simultaneous space and time reflections, $\psi^a_\gamma(x)\to\psi^a_\gamma(-x)$, leave the action invariant.

\medskip\noindent
{\bf 6. Discretization}

Next we formulate a regularized version of the functional integral \eqref{1A}. For this purpose we will use a lattice of space-time points. We recall that the action \eqref{30F} is invariant under $SO(4)$ and $SO(1,3)$ transformations and does not involve any metric. The regularization will therefore be valid simultaneously for a Minkowski and a euclidean theory.

Let us consider a four-dimensional hypercubic lattice with lattice distance $\Delta$. We distinguish between the ``even sublattice'' of points $y^\mu=\tilde y^\mu\Delta$, $\tilde y^\mu$ integer, $\Sigma_\mu\tilde y^\mu$ even, and the ``odd sublattice'' $z^\mu=\tilde z^\mu\Delta~,~\tilde z^\mu$ integer, $\Sigma_\mu\tilde z^\mu$ odd. The odd sublattice is considered as the fundamental lattice, and we associate to each position $z^\mu$ the $16$ (``real'') Grassmann variables $\psi^a_\gamma(z)$, or their complex counterpart $\varphi^a_\alpha(z)$. The functional measure is invariant under local $SO(4,{\mathbbm C})$ transformations since it can be written as a product of invariants of the type $A_+,A_-$ in eq. \eqref{32D}  and their complex conjugate for every $z$. It is also invariant under local $SU(2,{\mathbbm C})_L\times SU(2,{\mathbb C)})_R$ gauge transformations. 

For a finite number of lattice points the number of Grassmann variables is finite and the regularized functional integral is mathematically well defined. For example, this can be realized by a periodic lattice with $L$ lattice points on a torus in each ``direction'' $\mu$, such that the total number of lattice points is $N_L=L^4/2$. Alternatively, we could take some finite number of lattice points $L_t$ in some direction, without imposing a periodicity constraint. The continuum limit corresponds to $N_L\to \infty $ and is realized by keeping fixed $z^\mu$ with $\Delta\to 0$.

We write the action as a sum over local terms or Lagrangians ${\cal L}(y)$, 
\be\label{L1}
S=\tilde\alpha\sum_y{\cal L}(y)+c.c.
\ee
Here $y^\mu$ denotes a position on the even sublattice or ``dual lattice''. It has eight nearest neighbors on the fundamental lattice, with distance $\Delta$ from $y$. To each point $y$ we associate a ``cell'' of those eight points $\tilde x_j(\tilde y),j=1\dots 8$, with $\tilde z$-coordinates given by 
\be\label{V1}
\tilde z^\mu\big(\tilde x_j(\tilde y)\big)=\tilde y^\mu+V^\mu_j.
\ee
The eight vectors $V_j$ obey
\ba\label{V2}
V_1=(-1,0,0,0)&,&V_5=(0,0,0,1)\nn\\
V_2=(0,-1,0,0)&,&V_6=(0,0,1,0)\nn\\
V_3=(0,0,-1,0)&,&V_7=(0,1,0,0)\nn\\
V_4=(0,0,0,-1)&,&V_8=(1,0,0,0).
\ea
The distance between two neighboring $\tilde x_j$ is $\sqrt{2}\Delta$, and each point in the cell has six nearest neighbors. There is further an ``opposite point'' at distance $2\Delta$, with pairs of opposite points given by $(\tilde x_1,\tilde x_8),(\tilde x_2,\tilde x_7),(\tilde x_3,\tilde x_6),(\tilde x_4,\tilde x_5)$.

The Lagrangian ${\cal L}(y)$ is given by a sum of ``hyperloops''. A hyperloop is a product of an even number of Grassmann variables located at positions $\tilde x_j(\tilde y)$ within the cell at $\tilde y$. In accordance with eq. \eqref{A1} we will consider hyperloops with twelve spinors. In a certain sense the hyperloops are a four-dimensional generalization of the plaquettes in lattice gauge theories. 

\medskip\noindent
{\bf 7. Local $SO(4,{\mathbbm C}$ symmetry}

We want to preserve the local $SO(4,{\mathbbm C})$-symmetry for the lattice regularization of spinor gravity. We therefore employ hyperloops that are invariant under local $SO(4,{\mathbbm C})$ transformations. Local $SO(4,{\mathbbm C})$ symmetry can be implemented by constructing the hyperloops as products of invariant bilinears involving two spinors located at the same position $\tilde x_j(\tilde y)$,
\be\label{L4}
\tilde \h^k_\pm (\tilde x)=\varphi^a_\alpha(\tilde x)(C_\pm)_{\alpha\beta}(\tau_2\tau_k)^{ab}\varphi^b_\beta(\tilde x).
\ee
Since the local $SO(4,{\mathbbm C})$ transformations \eqref{A2} involve the same $\epsilon_{mn}(\tilde x)$ for both spinors the six bilinears $\tilde \h^k_\pm$ are all invariant. The three matrices $\tilde\tau_k=\tau_2\tau_k$ are symmetric, such that $C_\pm\otimes \tau_2\tau_k$ is antisymmetric, as required by the Pauli principle.

An $SO(4,{\mathbbm C})$ invariant hyperloop can be written as a product of six factors $\tilde \h(\tilde x_j\big(\tilde y)\big)$, with $\tilde x_j$ belonging to the hypercube $\tilde y$ and obeying eq. \eqref{V1}. We will take all six positions $\tilde x_{j_1}\dots \tilde x_{j_6}$ to be different. Furthermore, we will take three factors $\tilde \h_+$ and three factors $\tilde \h_-$ in order to realize the global symmetries of the continuum limit. The values of $k$ for the three factors $\tilde \h_+$ will be taken all different, and similar for the three factors $\tilde \h_-$. An invariant hyperloop is therefore fully specified by three positions $\{j_+\}=(j_1,j_2,j_3)$ for the bilinears $\tilde \h^1_+,\tilde \h^2_+$ and $\tilde \h^3_+$, and three positions $\{j_-\}=(j_4,j_5,j_6)$ for the bilinears $\tilde \h^1_-,\tilde \h^2_-$ and $\tilde \h^3_-$.

\medskip\noindent
{\bf 8. Lattice action}

The lattice action is a sum of local terms ${\cal L}(y)$ for all hypercubes $\tilde y$, where each ${\cal L}(y)$ is a combination of hyperloops. We consider a Lagrangian of the form 
\be\label{V3}
{\cal L}(y)=s\big\{{\cal F}_+^{1,2,8,7}(y){\cal F}^{3,4,6,5}_-(y)\big\}
\ee
with
\ba\label{V4}
&&{\cal F}^{abcd}_\pm=\frac{1}{24}\epsilon^{klm}\big[\tilde \h^k_\pm(\tilde x_a)\tilde \h^l_\pm(\tilde x_b)
\tilde \h^m_\pm(\tilde x_c)\nn\\
&&\quad +\tilde \h^k_\pm(\tilde x_b)\tilde \h^l_\pm(\tilde x_c)\tilde \h^m_\pm(\tilde x_d)+\tilde \h^k_\pm(\tilde x_c)
\tilde \h^l_\pm(\tilde x_d)\tilde \h^m_\pm(\tilde x_a)\nn\\
&&\quad +\tilde \h^k_\pm(\tilde x_d)\tilde \h^l_\pm(\tilde x_a)\tilde \h^m_\pm(\tilde x_b)\big].
\ea
The symbol $s$ denotes a symmetrization that will be discussed below. 

We observe that ${\cal F}^{1,2,8,7}_+$ is invariant under rotations by $\pi/2$ in the $z^0-z^1$-plane, corresponding to $\tilde x_1\to\tilde x_2~,~\tilde x_2\to\tilde x_8~,~\tilde x_8\to\tilde x_7~,~\tilde x_7\to\tilde x_1$. These rotations exchange the four terms cyclically in eq. \eqref{V4}, and we observe 
\be\label{V5}
{\cal F}^{abcd}_\pm={\cal F}^{bcda}_\pm={\cal F}^{cdab}_\pm={\cal F}^{dabc}_\pm.
\ee
A reflection $z^0\to -z^0$ exchanges $\tilde x_1\leftrightarrow \tilde x_8$, while all other $\tilde x_j$ remain invariant. (It also exchanges the positions of the cells by $y^0\to-y^0$. Since the action involves a sum over all positions $y$, it is sufficient to discuss rotations and reflections for the cell at $\tilde y=0$.) The reflection $\tilde x_1\leftrightarrow \tilde x_8$ maps 
\be\label{V6}
{\cal F}^{1,2,8,7}_+\to {\cal F}^{8,2,1,7}_+=-{\cal F}^{1,2,8,7}_+,
\ee
such that ${\cal L}(y)$ changes sign. Indeed, ${\cal F}^{abcd}_\pm$ is antisymmetric under the exchange of the two indices $a$ and $c$. This amounts to an exchange $k\leftrightarrow m$ for the first and third factor in eq. \eqref{V4}, whereas the second and fourth factor are mapped into each other, together with $k\leftrightarrow m$. The exchange $k\leftrightarrow m$ yields a minus sign due to the total antisymmetry of $\epsilon^{klm}$. 

Similarly, one finds antisymmetry in the second and fourth index of $\f$,
\be\label{VV1}
\f_\pm^{cbad}=\f^{adcb}_\pm =-\f^{abcd}_\pm,
\ee
implying that $\f^{1,2,8,7}_+$ is odd under the reflection $z^1\to -z^1$ which exchanges $\tilde x_2\leftrightarrow \tilde x_7$. Furthermore, we may consider a reflection on a diagonal in the $z^0-z^1$-plane, which exchanges simultaneously $\tilde x_1\leftrightarrow\tilde x_7$ and $\tilde x_2\leftrightarrow\tilde x_8$, resulting in $\f^{1,2,8,7}\to \f^{7,8,2,1}=-\f^{2,8,7,1}=-\f^{1,2,8,7}$. The same holds for the other diagonal reflection, $\tilde x_1\leftrightarrow\tilde x_2~,~\tilde x_7\leftrightarrow\tilde x_8$. Since $\f^{3,4,6,5}_-$ is invariant under reflections and rotations in the $z^0-z^1$-plane we conclude that ${\cal L}(y=0)$ and therefore also the action \eqref{L1} are invariant under $\pi/2$-rotations in the $z^0-z^1$-plane, while the action changes sign under the reflections $z^0\to -z^0~,~z^1\to -z^1~,~z^0\leftrightarrow z^1$ and $z^0\leftrightarrow -z^1$. These are the required symmetry properties for the continuum action. The same transformation properties hold for rotations and reflections in the $z^2-z^3$-plane. Now $\f^{1,2,8,7}_+$ is invariant, while $\f^{3,4,6,5}_-$ is even under $\pi/2$-rotations and odd under reflections.

The symmetrization $s$ in eq. \eqref{V3} sums over all six possibilities to place the factors $\tilde {\cal H}_+$ on the possible planes spanned by two coordinates $z^\mu$, e.g. $(0,1),(0,2)\dots (2,3)$. The signs of the hyperlinks are thereby chosen such that the six terms can be obtained from each other by $\pi/2$-rotations. We can write the symmetrization explicitly as
\ba\label{VV3}
{\cal L}(y)&=&\frac16\big \{\f^{1,2,8,7}_+\f^{3,4,6,5}_-+\f^{1,3,8,6}_+\f^{7,4,2,5}_-\nn\\
&&+\f^{1,4,8,5}_+\f^{3,7,6,2}_-+(\f_+\leftrightarrow \f_-)\big \}.
\ea

As a result, ${\cal L}(y)$ is invariant under $\pi/2$-rotations in all six planes spanned by two coordinates $z^\mu$. It is also odd under all four reflections of a single coordinate, $z^\mu\to -z^\mu$. For a ``diagonal reflection'' as $z^1\leftrightarrow z^2$ corresponding to $\tilde x_2\leftrightarrow \tilde x_3~,~\tilde x_6\leftrightarrow \tilde x_7$ we observe $\f^{1,2,8,7}_+\f^{3,4,6,5}_-\to\f^{1,3,8,6}_+\f^{2,4,7,5}_-=-\f^{1,3,8,6}\f^{7,4,2,5}$, such that the sum of the first two terms in eq. \eqref{VV3} changes sign. The third term is odd itself, and the three remaining terms obtained by exchanging $\varphi_+\leftrightarrow\varphi_-$ show the same transformation properties as the first three terms. Thus ${\cal L}(y)$ in eq. \eqref{VV3} is odd under this reflection, and the same holds for all twelve diagonal reflections of the type $z^\mu\leftrightarrow z^\nu$ or $z^\mu\leftrightarrow-z^\nu$. The discretized action \eqref{L1} shares with the continuum action the transformation properties with respect to $\pi/2$-rotations in all $z^\mu-z^\nu$-planes, as well as reflections of single $z^\mu$ or diagonal reflections. 

Finally, we note that the three components $\tilde \h^k_+$ in eq. \eqref{L4} transform as a three-component vector with respect to global $SU(2,{\mathbbm C})_L$ gauge transformations. Thus the contraction \eqref{V4} with the invariant tensor $\epsilon^{klm}$ yields a $SU(2,{\mathbbm C})_L$-singlet, and $\f^{abcd}_+$ is invariant under global $SU(2,{\mathbbm C})_L$ transformations. The lattice action is invariant under global $SU(2,{\mathbbm C})_L\times SU(2,{\mathbbm C})_R$ gauge transformations.

It is, however, not invariant under local gauge transformations of this kind. Local gauge transformations transform the factors $\tilde\h^k_\pm$ at different positions $\x j$ differently. If we would like to realize local $SU(2)$ gauge symmetry we would have to replace $(\tilde \tau_k)^{ab}$ in eq. \eqref{L4} by the invariant $\tilde\tau_0=\tau_2$. This is not compatible with local Lorentz symmetry. The $4\times 4$ matrices $C_\pm\otimes \tilde\tau_0$ are symmetric, such that $\tilde \h$ would vanish due to the Pauli principle. One could try to realize a local $U(1)$-symmetry by employing a different structure where only $\tilde \h^3_\pm$ appears. This is, however, not compatible with the required transformation properties of the lattice action with respect to reflections.

\medskip\noindent
{\bf 9. Lattice derivatives}

Lattice derivatives in the $z^\mu$-directions are defined, with $\hat\partial_\mu\widehat{=}\partial/\partial z^\mu$, as
\ba\label{VV4}
\hat\partial_0\varphi(y)&=&\frac{1}{2\Delta}\big(\varphi(\tilde x_8)-\varphi(\tilde x_1)\big),\nn\\
\hat\partial_1\varphi(y)&=&\frac{1}{2\Delta}\big(\varphi(\tilde x_7)-\varphi(\tilde x_2)\big),\nn\\
\hat\partial_2\varphi(y)&=&\frac{1}{2\Delta}\big(\varphi(\tilde x_6)-\varphi(\tilde x_3)\big),\nn\\
\hat\partial_3\varphi(y)&=&\frac{1}{2\Delta}\big(\varphi(\tilde x_5)-\varphi(\tilde x_4)\big). 
\ea
Here we have suppressed the spinor and flavor indices of $\varphi^a_\alpha$, and $\tilde x_j$ stands for $\tilde x_j(y)$. Note that we associate the lattice derivatives with positions $y$ on the dual lattice. To each position $y$ of a cell we can also associate ``average spinors''
\ba\label{VV5}
\bar\varphi_0(y)&=&\frac12\big(\varphi(\tilde x_1)+\varphi(\tilde x_8)\big)~,~\bar\varphi_1(y)=\frac12\big(\varphi(\tilde x_2)+\varphi(\tilde x_7)\big),\nn\\
\bar\varphi_2(y)&=&\frac12\big(\varphi(\tilde x_3)+\varphi(\tilde x_6)\big)~,~\bar\varphi_3(y)=\frac12\big(\varphi(\tilde x_4)+\varphi(\tilde x_5)\big),\nn\\
\ea
and we write for each cell $y$
\ba\label{VV6}
\varphi(\x j)=\sigma^\mu_j\bar\varphi_\mu+ V^\mu_j\Delta\hat\partial_\mu\varphi,
\ea
with $\sigma^\mu_j=(V^\mu_j)^2$.

We next express ${\cal L}(y)$ in terms of the averages $\bar\varphi$ and lattice derivatives $\hat\partial_\mu\varphi$. We use
\ba\label{VV7}
\tilde{\cal H}^k_\pm(\tilde x_j)=\sigma^\mu_j\bar{\cal H}^k_{\pm\mu}(y)+2\Delta V^\mu_j
{\cal D}^{k}_{\pm\mu}(y)+\Delta^2\sigma^\mu_j{\cal G}^{k}_{\pm\mu}(y),
\ea
where $\bar{\cal H}_\mu(y)$ obtains from $\tilde {\cal H}(\tilde x_j)$ by the replacement $\varphi(\tilde x_j)\to\bar\varphi_\mu(y)$. We also define (no sum over $\mu$ here)
\be\label{W8} 
\D^k_{\pm\mu}=(\bar\varphi_\mu)^a_\alpha(C_\pm)_{\alpha\beta}(\tau_2\tau_k)^{ab}\hat\partial_\mu\varphi^b_\beta
\ee
as well as
\be\label{VV9}
\g^k_{\pm\mu}=\hat\partial_\mu\varphi^a_\alpha(C_\pm)_{\alpha\beta}(\tau_2\tau_k)^{ab}\hat\partial_\mu\varphi^b_\beta,
\ee
and
\ba\label{VV11}
\hat\h^k_{\pm \mu}=\bar \h^k_{\pm \mu}+\Delta^2\tilde \g^k_{\pm\mu}~,~\h^k_{\pm \mu\nu}=\frac12(\hat \h^k_{\pm \mu}+
\hat\h^k_{\pm \nu}).
\ea
Expanding $\f^{1,2,8,7}_+$ in powers of $\Delta$ one finds 
\be\label{VV16}
\f^{1,2,8,7}_+=\frac{2\Delta^2}{3}\epsilon^{klm}\h^k_{+01}
(\D^l_{+0} \D^m_{+1}-\D^l_{+1}\D^m_{+0}).
\ee

Since the four points $\tilde x_1,\tilde x_2,\tilde x_8$ and $\tilde x_7$ are all in the $z^0-z^1$-plane of the cell we may switch notation and denote 
\be\label{VV17}
\f^\pm_{01}=-\f^\pm_{10}=\f^{1,2,8,7}_\pm.
\ee
Similarly, one defines the other $\f^\pm_{\mu\nu}=-\f^\pm_{\nu\mu}$
\ba\label{VV18}
\f^\pm_{02}=\f_\pm^{1,3,8,6}~,~\f^\pm_{03}=\f^{1,4,8,5}_\pm,
\ea
and 
\ba\label{VV19}
\f^\pm_{12}=\f^{3,7,6,2}_\pm~,~\f^\pm_{13}=\f^{2,4,7,5}~,~\f^\pm_{23}=\f^{3,4,6,5}_\pm.
\ea
The relations similar to eq. \eqref{VV16} can then be summarized as
\be\label{VV20}
\f^\pm_{\mu\nu}=\frac{2\Delta^2}{3}\epsilon^{klm}\h^k_{\pm\mu\nu}
(\D^l_{\pm\mu}\D^m_{\pm\nu}-\D^l_{\pm\nu}\D^m_{\pm \mu}).
\ee
In this notation we find the intuitive expression 
\be\label{VV21}
{\cal L}(y)=\frac{1}{24}\epsilon^{\mu_1\mu_2\mu_3\mu_4}\f^+_{\mu_1\mu_2}\f^-_{\mu_3\mu_4}.
\ee
This structure is very similar to the continuum action \eqref{32F}. 

\medskip\noindent
{\bf 10. Continuum limit}

The continuum limit obtains formally as $\Delta\to 0$ at fixed $y^\mu$. In this limit we can replace $\bar\varphi_\mu(y)$ by $\varphi(y)$ and the lattice derivative $\hat\partial_\mu$ becomes the continuum derivative $\partial_\mu=\partial/\partial y^\mu$. This results in 
\be\label{VV22}
\D^k_{\pm \mu}(y)\to \varphi^a_\alpha (y)(C_\pm)_{\alpha\beta}(\tau_2\tau_k)^{ab}\partial_\mu\varphi^b_\beta(y),
\ee
and 
\be\label{VV23}
\h^k_{\pm\mu\nu}(y)\to\varphi^a_\alpha(y)(C_\pm)_{\alpha\beta}(\tau_2\tau_k)^{ab}\varphi^b_\beta(y)=H^\pm_k (y).
\ee
Insertion of the continuum limit \eqref{VV22}, \eqref{VV23} into eq. \eqref{VV20} yields the simple result 
\be\label{VV34}
\f^+_{\mu\nu}=16i\Delta^2 F^+_{\mu\nu}~,~\f^-_{\mu\nu}=-16i\Delta^2F^-_{\mu\nu}.
\ee

We infer the continuum limit of ${\cal L}(y)$, 
\be\label{VV31}
{\cal L}(y)\to\frac{32}{3}\Delta^4\epsilon^{\mu_1\mu_2\mu_3\mu_4}
F^+_{\mu_1\mu_2}F^-_{\mu_3\mu_4}.
\ee
For a computation of the action we have to convert the sum over cells $\Sigma_y$ into an integral 
\be\label{VV37}
\Delta^4\sum_y=\frac12\int_y,
\ee
where the factor $1/2$ accounts for the fact that the positions $y$ of the cells are placed only on the even sublattice of a hypercubic lattice. The factor $\Delta^4$ cancels in $\Sigma_y{\cal L}(y)$ and the action is independent of the lattice distance $\Delta$
\be\label{VV38}
S=\frac{16}{3}\tilde\alpha\int_y\epsilon^{\mu_1\mu_2\mu_3\mu_4}F^+_{\mu_1\mu_2}
F^-_{\mu_3\mu_4}+c.c
\ee
This coincides with the continuum action \eqref{30F}, provided we choose $\tilde\alpha=3\alpha/16$. 

\medskip\noindent
{\bf 11. Emergent geometry}

By now we have achieved the main task of this letter, namely the construction of a lattice action that fulfills the four criteria mentioned in the introduction. (The geometrical origin of the diffeomorphism symmetry of the continuum action will be discussed elsewhere.) We sketch here only how geometry can arise in this setting. We can introduce a ``global vierbein bilinear''
\ba\label{BB3}
\tilde E^m_\mu=\varphi^aC\gamma^m_M\partial_\mu\varphi^b V^{ab}=-\partial_\mu\varphi^a C\gamma^m_M\varphi^b V^{ab}.
\ea
Here $V^{ab}$ is a suitable flavor matrix and we choose $C=C_1$ for symmetric $V$, and $C=C_2$ if $V$ is antisymmetric.

With respect to diffeomorphisms $\tilde E^m_\mu$ transforms as a covariant vector. It also transforms as a vector with respect to global $SO(4,{\mathbbm C})$ transformations. This can be seen most easily by switching to the ``euclidean vierbein bilinear''
\be\label{A19}
\tilde E^{(E)m}_\mu=\varphi V C\gamma^m\partial_\mu\varphi 
\ee
by multiplication with an appropriate factor $i$,
\be\label{A20}
\tilde E^{(E)0}_\mu=i\tilde E^0_\mu~,~\tilde E^{(E)k}_\mu=\tilde E^k_\mu.
\ee
Using Minkowski transformation parameters  $\epsilon^{(M)}\tmn$ one obtains for real $\epsilon^{(M)}\tmn$ and $\epsilon^{(M)n}_m=\eta^{np}\epsilon^{(M)}_{mp}$ the standard homogeneous transformation property of a Lorentz vector 
\be\label{A21}
\delta\tilde E^{m}_\mu=-\tilde E^{n}_\mu
\epsilon_n^{(M)m}.
\ee
We conclude that $\tilde E^{m}_\mu$ has almost the transformation properties of the vierbein in general relativity with a Minkowski signature. In distinction to the usual vierbein, a local Lorentz transformation of the object \eqref{A19} also generates an inhomogeneous term. For the dimensionless vierbein
\be\label{76A}
\tilde e^m_\mu=\Delta\tilde E^m_\mu
\ee
this inhomogeneous term vanishes in the continuum limit $\Delta\to 0$. 

We will be interested in a situation where
\be\label{50M}
\kl \tilde e^{m}_\mu\kr=\kl(\tilde e^{m}_\mu)^*\kr= e^m_\mu.
\ee
We associate the ``background vierbein'' $e^m_\mu$ with the vierbein that defines the geometry of spacetime via the metric
\be\label{B5-1}
g_{\mu\nu}=e^m_\mu e^n_\nu\eta_{mn}.
\ee
Thus geometry emerges from properties of expectation values of composite bosonic observables. 

One may ask if it is possible to write the action \eqref{30F} in the intuitive form 
\be\label{A18}
S=\alpha\int d^4 x W \det (\tilde E^m_\mu)+c.c.,
\ee
with $\tilde E^m_\mu$ given by eq. \eqref{BB3} and interpreted as a matrix with first index $\mu$ and second index $m$. The invariant $W$ has to involve two Weyl spinors $\varphi_+$ and two Weyl spinors $\varphi_-$. It should be a singlet with respect to $SO(4,{\mathbbm C})$-transformations. The invariance of the action \eqref{A18} under diffeomorphisms and $SO(4,{\mathbbm C})$ transformations would be particularly transparent in this language. With respect to diffeomorphisms the determinant $\tilde E=\det(\tilde E^{m}_\mu)$ has the same transformation properties as the determinant of the vierbein in general relativity. The latter equals the usual volume factor $\sqrt{g}=|\det (g_{\mu\nu})|^{1/2}$, and we recover the general coordinate invariance of the action \eqref{A18}. Since $W$ contains no derivatives it transforms as a scalar. For $\kl W\det \tilde E^m_\mu\kr\approx \kl W\kr\det e^m_\mu/\Delta^4$ the action \eqref{A18} would describe a cosmological constant $\sim \kl W\kr/\Delta^4$. One can show, however, that no invariant $W$ exists which is compatible with the symmetries of the action. 

\medskip\noindent
{\bf 12. Conclusions}

We have constructed a lattice regularized functional integral for fermions with local Lorentz symmetry. The continuum limit of the action exhibits invariance under general coordinate transformations. We postpone a geometrical discussion of the lattice origins of diffeomorphisms to a separate publication. At this point we only emphasize the striking fact that for a fixed definition of lattice derivatives the lattice distance $\Delta$ drops out in the continuum limit. This property requires that the continuum limit contains $d$ derivatives for a $d$-dimensional theory. It is not shared by general lattice models as, for example, standard lattice gauge theories. 

The symmetry properties of our model suggest that it can be used as a promising starting point for quantum gravity. We have only sketched the way to geometry. Much remains to be done before the effective action for a composite metric can be computed explicitly. For our regularized model this issue is, at least, well defined. However, only an explicit calculation can settle the issue if diffeomorphism invariant terms involving length scales, as a cosmological constant or Einstein's curvature scalar multiplied by the Planck mass, can be generated by fluctuations. The classical continuum action \eqref{30F} is dilatation symmetric - the only coupling $\alpha$ is dimensionless. If the effective action for the graviton preserves dilatation symmetry no dimensional couplings can be present. In this case one would expect gravitational invariants involving two powers of the curvature tensor, as $R_{\mu\nu\rho\sigma}R^{\mu\nu\rho\sigma}$, $R_{\mu\nu}R^{\mu\nu}$ or $R^2$. Also composite scalar fields may play a role, such that terms $\sim \xi R$ can induce an Einstein-Hilbert term in the effective action by spontaneous dilatation symmetry breaking through an expectation value of $\xi$ \cite{CWDil}. As an alternative, an explicit mass scale could be generated by running couplings, which constitute a dilatation anomaly through quantum fluctuations. 

Our lattice construction admits straightforward generalizations. The number of dimensions (even $d$) and flavors is not restricted. Local gauge symmetries can be realized by replacing $\h^k_\pm$ in eq. \eqref{L4} by a local gauge singlet. For a formulation of a lattice gauge theory purely in terms of fermions the local Lorentz symmetry is not necessary - it is sufficient that the lattice action exhibits global Lorentz symmetry. In the present setting we can construct a gauge theory with local $SU(2)_L\times SU(2)_R$ symmetry by replacing in eq. \eqref{L4} the tensor $C_\pm\otimes\tilde\tau_k$ by $C_\pm\tilde\tau_k\otimes\tilde\tau_0,\tilde\tau_0=\tau_2$. For a different number of flavors or dimensions we can also construct lattice theories realizing simultaneously local Lorentz and gauge symmetries. It is sufficient that at least three different singlets (with respect to both Lorentz and gauge transformations) exist in the antisymmetric product of two spinors. This allows for the construction of $\f_\pm$ in eq. \eqref{V4}  with the required symmetry properties under rotations and reflections. (For more than three singlets $\epsilon^{klm}$ can be replaced by a more general totally antisymmetric three index tensor. One could also use four factors of $\h$ contracted by a suitable tensor, such that every point in a given surface of the cell carries an invariant bilinear.) If the complexified group $SO(4,{\mathbbm C})$ plays no role one can also use instead of $\h^k_\pm$ invariants involving products of $\varphi$ and $\varphi^*$. Beyond the use for a regularized model for quantum gravity our lattice construction can be employed for formulating a wide class of theories with local gauge invariance purely in terms of fermions.

\end{document}